\documentclass{bulsao}
\usepackage{graphicx}
\usepackage{latexsym}
\usepackage{longtable}

\begin{document}

\title{Speckle Interferometry of Nearby Multiple Stars. IV. Measurements in 2004 and New Orbits}

\author{I.I.~Balega$^1$, Yu.Yu.~Balega$^1$, A.F.~Maksimov$^1$, E.V.~Malogolovets$^1$,
D.A.~Rastegaev$^1$, Z.U.~Shkhagosheva$^1$, G.~Weigelt$^2$}

\institute{Special Astrophysical Observatory, RAS, Nizhnii Arkhyz,
Karachai-Cherkessian Republic, 357147 Russia \and Max-Planck Institut
f\"{u}r Radioastronomie, Bonn, Germany}

\offprints{Yu.Yu.~Balega, \email{balega@sao.ru}}

\date{received: August 27, 2007/revised: October 3, 2007}

\titlerunning{SPECKLE INTERFEROMETRY OF NEARBY MULTIPLE STARS. IV.}
\authorrunning{Balega et al.}

\onecolumn
\abstract{
The results of speckle interferometric observations of 104 binary and 6
triple stars performed at the BTA 6~m telescope in 2004 October are
presented. Nearby low-mass stars are mostly observed for the
program, among which 59 there are new binaries recently discovered by the Hipparcos
astrometric satellite. Concurrently with the diffraction-limited
position measurements we obtained 154 brightness ratio measurements
of binary and multiple star components in different bands of the visible
spectrum. New, first-resolved binaries are the symbiotic star CH\,Cyg
with a weak companion at 0.043$''$ separation and the pair of red dwarfs,
GJ\,913\,=\,HIP\,118212. In addition, we derived the orbital parameters
for two interferometric systems: the CN-giant pair HD\,210211\,=\,HIP\,109281
(P=10.7 yr) and the G2V-K2V binary GJ\,9830\,=\,HIP\,116259
(P=15.7 yr).
}

\maketitle

\section{INTRODUCTION}
This is the fourth paper in the series of publications with the data on speckle
interferometry observations of binary and multiple stars performed with the BTA
6~m telescope of the Special Astrophysical Observatory of the Russian Academy of Sciences
 using a new detector
system based on a 3-stage image intensifier and a fast CCD (\cite{maks2003:Balega_n}).
The main objects of the program are nearby low-mass stars with a considerable,
of the order of 10$\degr$/yr, relative motion of the components, which makes
them good new candidates for the calculation of visible orbits. Around half
of these stars are new binaries discovered by the Hipparcos astrometric
satellite (\cite{eca:Balega_n}). The regular speckle interferometric observations of
new Hipparcos binaries have been carried out at the BTA telescope since 1998
(\cite{bag2002:Balega_n}; \cite{bag2004:Balega_n}; \cite{bag2006a:Balega_n}). In addition, some
early-type systems that
are interesting for interferometric monitoring were included in the program.
In particular, the Orion Trapezium members were observed in the visible range
to reveal the relative motion of the components.

\section{OBSERVATIONS AND RESULTS}

The measurements are derived from speckle interferometry (\cite{laber:Balega_n})
observations taken at the BTA 6~m telescope of the Special Astrophysical
Observatory during the period October
23 through November 1, 2004. During the observing period, the
seeing changed between 1$''$ and 5$''$. On October 25/26, the seeing was
0.8$''$--1$''$. Note that even in the nights of poor seeing, speckle
interferometry allowed us to perform speckle measurements of bright stars
with a diffraction-limited angular resolution.

The instrumentation, observing procedure, data
reduction, and calibration have already been described in the previous papers
of this series (\cite{bag2002:Balega_n}; \cite{bag2004:Balega_n}; \cite{bag2006a:Balega_n}). The
high sensitivity of the detector allows us to measure stars up to the 15th magnitude with a
diffraction-limited resolution.

In this paper the results of 181 measurements of the relative positions of
104 binary stars (Table~\ref{tab1:Balega_n}) and single measurements for 6 triple
stars (Table~\ref{tab2:Balega_n}) are presented. For each system the tables give
four identifier numbers (the Hipparcos Catalog number, the name or the number from
other catalogs, the discoverer designation, and the Washington Double Star
Catalog coordinates, J2000.0).
The identifier numbers are followed by the observation date as a fraction
of the Besselian year, the measured position
angle $\theta$ in degrees and its error $\sigma_{\theta}$, the measured
angular distance $\rho$ in milliarcseconds (mas) and its error $\sigma_{\rho}$,
the observed magnitudes difference $\Delta m$ and its uncertainty
$\sigma_{\Delta m}$, the center wavelength $\lambda$ of the filter used to
make the observation (nm), and the FWHM of the filter passband
$\Delta\lambda$. For triple stars, Table~\ref{tab2:Balega_n}
presents also the designations of the subsystems.
The measured distances between the components of the
systems range from 23~mas for $\theta^{1}$\,Ori\,C to 1622~mas for
HIP\,103810. The separation accuracy depends on many parameters;
first of all, on the atmospheric conditions. For the majority of measurements,
it is equal to 2--3~mas; however, for the most wide pairs with a separation
of $>1''$, the error may reach 6--8~mas.  The errors of the position angle
measurements are 0.3$\degr$--1.0$\degr$. Comments on the
measurements of individual stars are given in the next section.

It is known that in speckle interferometry, the ensemble average modulus of
the Fourier transform of a series of speckle images defines the position angles of
binary stars with a $\pm$180$\degr$ ambiguity. To avoid this uncertainty, it is
necessary to reconstruct not only the modulus but also the phase of the
observed source (\cite{weigelt:Balega_n}; \cite{lohm:Balega_n}). This requires a large number of
additional computations. In binary star speckle interferometry, we solve the problem of
position angles using a simple approach proposed by Walker (\cite{walker:Balega_n}).
In this method we calculate the modulus of the Fourier
transform of the product of the speckle interferograms and an exponential in
addition to the measured modulus of the Fourier transform. From the
measurements of the two moduli the location of the complex zeros of the analytical
continuation of
the Fourier transform of the unknown image can be found and the true image of
a binary reconstructed. Problems arise when the components of a binary have
similar magnitudes or when the differential speckle photometry of the pair is
seriously noise-limited. The $\theta$ measurements with the $\pm$180$\degr$ ambiguity are
marked with asterisks in Table~\ref{tab1:Balega_n}.

\begin{longtable}{l|l|l|c|c|c|c|c|c|c|c|l}
\caption{Double star measurements}
\label{tab1:Balega_n}\\
\hline
HIP     & Name/         & Discoverer      & Coord. & Epoch   & $\theta$, deg & $\sigma_{\theta}$ & $\rho$, & $\sigma_{\rho}$, & $\Delta$m & $\sigma_{\Delta m}$ & $\lambda/\Delta\lambda$, \\
No.     & Catalog No.   & designation     & 2000.0 & 2004.0+ &             &                     & mas    & mas             &           &                     &   nm                    \\
\hline
\endfirsthead
\caption{(Contd.)} \\
\hline
HIP     & Name/         & Discoverer      & Coord. & Epoch   & $\theta$, deg & $\sigma_{\theta}$ & $\rho$, & $\sigma_{\rho}$, & $\Delta$m & $\sigma_{\Delta m}$ & $\lambda/\Delta\lambda$, \\
No.     & Catalog No.   & designation     & 2000.0 & 2004.0+ &             &                     & mas    & mas             &           &                     &   nm                    \\
\hline
\endhead
\hline
\endfoot
\endlastfoot
68      & BD+16 5027    & BAG 18        & 00008+1659 & .8318 & 22.3       & 0.3   & 560    & 2  & 2.68  & 0.04 & 800/110 \\
201     & HD 225000     & HDS 2         & 00026+1841 & .8372 & 123.3$\ast$& 1.0   & 80     & 2  & 2.34  & 0.05 & 545/30  \\
689     & HD 375        & HDS 17        & 00085+3456 & .8237 & 347.9      & 0.5   & 64     & 2  & 0.20  & 0.04 & 600/30  \\
823     &               & HDS 23        & 00101+3825 & .8237 & 91.7$\ast$ & 1.18  & 72     & 2  & 0.00  & 0.21 & 800/110 \\
1055    & BD+19 20      & HDS 29        & 00132+2023 & .8238 & 169.0      & 0.3   & 664    & 2  & 1.11  & 0.06 & 800/110 \\
1987    & HD 2057       & HDS 56        & 00252+4803 & .8265 & 149.0$\ast$& 1.1   & 238    & 5  &       &      & 545/30  \\
2532    & HD 2893       & HDS 71        & 00321-1218 & .8342 & 153.7$\ast$& 0.4   & 292    & 2  & 0.41  & 0.10 & 545/30  \\
3361    & BD+12 81      & HDS 93        & 00428+1249 & .8211 & 71.3       & 0.3   & 247    & 2  & 1.44  & 0.03 & 600/30  \\
3669    & BD+42 170     & HDS 102       & 00469+4339 & .8320 & 125.6      & 0.6   & 152    & 2  & 1.06  & 0.03 & 800/110 \\
4267    & ADS 746       & STT 20        & 00546+1911 & .8212 & 188.0      & 0.4   & 540    & 3  & 1.05  & 0.05 & 545/30  \\
        &               &               &            & .8212 & 188.0      & 0.4   & 539    & 3  & 1.05  & 0.05 & 600/30  \\
    &               &               &            & .8212 & 187.7      & 0.4   & 540    & 3  & 0.99  & 0.05 & 850/75  \\
    &               &               &            & .8237 & 188.5      & 0.4   & 544    & 3  & 1.00  & 0.05 & 545/30  \\
    &               &               &            & .8237 & 188.5      & 0.4   & 542    & 3  & 0.98  & 0.05 & 600/30  \\
    &               &               &            & .8237 & 188.4      & 0.4   & 540    & 3  & 0.80  & 0.05 & 800/110 \\
    &               &               &            & .8237 & 188.3      & 0.4   & 539    & 3  & 0.83  & 0.05 & 850/75  \\
4809    & HD 6009       & HDS 134       & 01017+2518 & .8154 & 318.8      & 0.3   & 89     & 2  & 0.21  & 0.04 & 600/30  \\
4849    & GJ 3071       & HDS 135       & 01024+0504 & .8155 & 135.3      & 0.3   & 275    & 2  & 1.68  & 0.03 & 600/30  \\
5531    & HD 6840       & HDS 155       & 01108+6747 & .8155 & 159.6      & 0.3   & 116    & 2  & 0.71  & 0.02 & 545/30  \\
5674    & HD 7169       & HDS 160       & 01129+5136 & .8373 & 54.6       & 0.4   & 181    & 2  & 2.00  & 0.02 & 545/30  \\
5952    & HD 7640       & HDS 169       & 01166+1831 & .8238 & 247.9      & 0.4   & 639    & 4  & 3.32  & 0.17 & 600/30  \\
6060    & ADS 1040      & STF 102       & 01178+4901 & .8265 & 273.3      & 0.4   & 475    & 3  & 0.83  & 0.13 & 545/30  \\
    &               &               &            & .8265 & 273.9      & 0.6   & 474    & 5  & 0.47  & 0.37 & 800/110 \\
7338    &               & HDS 211       & 01345+7804 & .8156 & 245.2      & 0.8   & 279    & 4  & 2.17  & 0.06 & 800/110 \\
7397    &               & HDS 213       & 01463+4059 & .8239 & 202.0$\ast$& 0.4   & 80     & 2  & 0.00  & 0.13 & 545/30  \\
10022   & HD 13102      & COU 1067      & 02090+3540 & .8374 & 30.2       & 0.5   & 196    & 2  & 0.00  & 0.25 & 545/30  \\
10414   &               & HDS 297       & 02142+0909 & .8212 & 38.9       & 0.5   & 383    & 3  & 1.32  & 0.05 & 800/110 \\
10660   & HD 13865      & HDS 302       & 02172+5838 & .8374 & 243.7      & 0.5   & 392    & 3  & 2.73  & 0.09 & 545/30  \\
11253   & HD 14874      & HDS 314       & 02249+3039 & .8239 & 276.9      & 0.3   & 372    & 2  & 2.68  & 0.05 & 545/30  \\
11352   & HD 15013      & HDS 318       & 02262+3428 & .8157 & 185.4      & 0.4   & 124    & 2  & 0.00  & 0.17 & 600/30  \\
11474   & HR 719        & KUI 8         & 02280+0158 & .8213 & 37.3       & 0.3   & 502    & 3  & 0.25  & 0.04 & 545/30  \\
    &               &               &            & .8213 & 37.2       & 0.3   & 501    & 3  & 0.23  & 0.02 & 545/30  \\
    &               &               &            & .8213 & 37.4       & 0.3   & 505    & 3  & 0.18  & 0.02 & 800/110 \\
    &               &               &            & .8239 & 37.6       & 0.3   & 504    & 3  & 0.25  & 0.25 & 545/30  \\
    &               &               &            & .8239 & 37.6       & 0.3   & 504    & 3  & 0.23  & 0.04 & 600/30  \\
    &               &               &            & .8240 & 37.5       & 0.3   & 505    & 3  & 0.00  & 0.18 & 800/110 \\
12495   & ADS 2018 Aa   & CHR 208       & 02407+6117 & .8264 &269.2$\ast$ & 0.7   & 289    & 3  &       &      & 545/30  \\
12552   & HD 16656      & COU 1511      & 02415+4053 & .8374 & 65.0       & 0.7   & 135    & 2  & 0.65  & 0.06 & 545/30 \\
13308   & ADS 2165      & BU 1316       & 02512+6023 & .8264 & 297.7$\ast$& 0.5   & 317    & 3  &       &      & 545/30  \\
14075   & HD 18774      & HDS 385       & 03014+0615 & .8157 & 166.2      & 0.4   & 162    & 2  & 0.00  & 0.17 & 800/110 \\
14230   & HD 18940      & HDS 389       & 03035+2304 & .8157 & 23.1       & 0.5   & 76     & 2  & 1.73  & 0.06 & 545/30  \\
    &               &               &            & .8266 & 23.8       & 1.7   & 73     & 3  &       &      & 545/30  \\
14669   & GJ 125        & HDS 404       & 03095+4544 & .8213 & 240.5      & 0.3   & 83     & 2  & 1.59  & 0.05 & 800/110 \\
14864   & GJ 3206       & HDS 407       & 03119+6131 & .8214 & 156.3      & 0.3   & 600    & 2  & 1.51  & 0.03 & 800/110 \\
14929   & HD 19895      & HDS 408       & 03125+1857 & .8158 & 122.0      & 1.6   & 26     & 2  & 0.00  & 0.36 & 545/30  \\
15309   & ADS 2436      & STT 52        & 03175+6540 & .8156 & 59.1       & 0.4   & 485    & 2  & 0.45  & 0.05 & 545/30  \\
    &               &               &            & .8156 & 59.6       & 0.4   & 484    & 2  & 0.48  & 0.03 & 600/30  \\
    &               &               &            & .8156 & 59.8       & 0.4   & 484    & 2  & 0.33  & 0.09 & 800/110 \\
15737   & 63 Ari        & HDS 423       & 03228+2045 & .8267 & 292.9      & 0.5   & 416    & 4  & 3.36  & 0.12 & 700/30  \\
16025   & HD 21183      & HDS 430       & 03264+3520 & .8158 & 244.2      & 0.4   & 279    & 2  & 1.76  & 0.03 & 545/30  \\
18089   & 31 Tau        & KUI 15        & 03519+0633 & .8159 & 207.0      & 0.4   & 757    & 2  & 0.31  & 0.05 & 545/30  \\
    &               &               &            & .8159 & 207.2      & 0.4   & 757    & 2  & 0.37  & 0.09 & 600/30  \\
    &               &               &            & .8159 & 207.2      & 0.4   & 758    & 2  & 0.47  & 0.06 & 800/110 \\
    &               &               &            & .8267 & 207.3      & 0.4   & 758    & 2  &       &      & 545/30  \\
    &               &               &            & .8267 & 207.5      & 0.4   & 758    & 2  &       &      & 600/30  \\
    &               &               &            & .8267 & 207.4      & 0.4   & 758    & 2  &       &      & 700/30  \\
    &               &               &            & .8267 & 207.6      & 0.5   & 758    & 4  &       &      & 850/75  \\
    &               &               &            & .8268 & 207.6      & 0.4   & 758    & 2  &       &      & 800/110 \\
18370   & HD 24431      & HDS 494       & 03556+5238 & .8266 & 177.7$\ast$& 0.4   & 723    & 5  &       &      & 545/30  \\
18856   & BD+06 620     & HDS 510       & 04025+0638 & .8214 & 150.3      & 0.7   & 77     & 2  & 0.28  & 0.05 & 800/110 \\
    & HD 25811      & BAG 4         & 04063+1952 & .8158 & 229.0      & 0.9   & 74     & 2  & 0.24  & 0.27 & 545/30  \\
19206   & HD 26040      & HDS 521       & 040700-1000& .8240 & 350.5      & 0.3   & 234    & 2  & 1.39  & 0.02 & 545/30  \\
19270   & SZ Cam        & CHR 209       & 04078+6220 & .8216 & 115.6      & 0.3   & 75     & 2  & 0.95  & 0.02 & 545/30  \\
    &               &               &            & .8266 & 115.9      & 1.9   & 74     & 3  &       &      & 545/30  \\
19472   & HD 285465     & HEI 35        & 04102+1722 & .8241 & 343.4      & 0.3   & 323    & 2   & 1.29 & 0.04 & 600/30  \\
19591   & HD 284163     & CHR 14        & 04119+2338 & .8214 & 5.7        & 0.3   & 280    & 2   & 1.24 & 0.02 & 800/110 \\
20553   & HD 27836      & HDS 564       & 04242+1445 & .8159 & 247.2      & 0.4   & 302    & 2   & 2.24 & 0.04 & 800/110 \\
20777   & DF Tau        & THB 1         & 04270+2542 & .8215 & 247.1      & 0.5   & 108    & 2   & 0.60 & 0.03 & 800/110 \\
20895   & HD 283646     & HDS 576       & 04287+2613 & .8241 & 140.4      & 0.5   & 147    & 2   & 0.18 & 0.14 & 800/110 \\
21280   & HD 285931     & CHR 17        & 04340+1510 & .8160 & 271.1      & 0.4   & 192    & 2   & 1.05 & 0.04 & 800/110 \\
21762   & HD 29608      & CHR 154       & 04404+1631 & .8242 & 44.2       & 0.3   & 226    & 2   & 1.35 & 0.03 & 800/110 \\
21881   & 94 Tau        & MCA 16        & 04422+2257 & .8242 & 44.0       & 0.3   & 303    & 2   & 2.48 & 0.02 & 545/30  \\
22550   & ADS 3475      & BU 883        & 04512+1104 & .8241 & 55.6       & 0.3   & 96     & 2   & 0.00 & 0.11 & 545/30  \\
    &               &               &            & .8241 & 56.0       & 0.4   & 97     & 2   & 0.33 & 0.05 & 800/110 \\
23699   & HD 32641      & STT 97        & 05056+2304 & .8161 & 149.5      & 0.4   & 356    & 3   & 1.34 & 0.08 & 545/30  \\
    &               &               &            & .8161 & 149.4      & 0.4   & 358    & 3   & 1.32 & 0.04 & 800/110 \\
    &               &               &            & .8162 & 149.4      & 0.5   & 357    & 3   & 1.40 & 0.09 & 600/30  \\
    &               &               &            & .8216 & 149.6      & 0.4   & 354    & 3   & 1.16 & 0.02 & 545/30  \\
    &               &               &            & .8216 & 149.5      & 0.4   & 354    & 3   & 1.20 & 0.02 & 600/30  \\
    &               &               &            & .8216 & 149.4      & 0.4   & 357    & 3   & 1.15 & 0.02 & 850/75  \\
    &               &               &            & .8244 & 149.2      & 0.4   & 355    & 3   & 1.47 & 0.02 & 545/30  \\
    &               &               &            & .8244 & 149.1      & 0.4   & 356    & 3   & 1.48 & 0.02 & 600/30  \\
    &               &               &            & .8245 & 149.1      & 0.4   & 356    & 3   & 1.39 & 0.02 & 800/110 \\
23772   & HD 240622     & HDS 666       & 05066+2630 & .8268 & 207.4$\ast$& 1.6   & 169    & 5   &      &      & 800/110 \\
25499   & 115 Tau       & MCA 19        & 05272+1758 & .8269 & 94.7$\ast$ & 0.4   & 88     & 2   & 0.95 & 0.02 & 545/30  \\
25565   & IU Aur        & HDS 721       & 05279+3447 & .8268 & 49.4$\ast$ & 1.4   & 141    & 4   &      &      & 545/30  \\
25733   & ADS 4072      & HU 217        & 05297+3523 & .8268 & 253.9      & 0.4   & 604    & 3   & 1.70 & 0.11 & 545/30  \\
26220   &$\theta^{1}$ Ori A & PTR 1     & 05353-0523 & .8161 & 0.3        & 1.6   & 203    & 2   & 2.66 & 0.13 & 800/110 \\
    &               &               &            & .8215 & 0.9        & 0.8   & 205    & 3   & 4.14 & 0.16 & 545/30  \\
26221   &$\theta^{1}$ Ori C & WGT 1     & 05353-0523 & .8216 & 189.8      & 2.4   & 23     & 2   & 1.06 & 0.11 & 545/30  \\
29269   & HD 39861      & HDS 841       & 06102+8131 & .8270 & 197.5      & 0.5   & 654    & 5   &      &      & 800/110 \\
30272   & ADS 4950 AB   & STF 881       & 06221+5922 & .8270 & 143.1      & 0.3   & 657    & 2   &      &      & 700/30  \\
    &               &               &            & .8270 & 143.7      & 0.3   & 657    & 3   &      &      & 800/110 \\
    &               &               &            & .8271 & 143.2      & 0.3   & 660    & 2   &      &      & 545/30  \\
30920   & GJ 234        &  B 2601       & 06294-0249 & .8217 & 37.1     & 0.3   & 1359   & 3  & 2.77$\ast$& 0.03 & 800/110 \\
32132   & BD+40 1685    & HDS 930       & 06426+3955 & .8244 & 20.4$\ast$ & 0.6   & 88     & 2   & 0.00 & 0.18 & 545/30  \\
32313   & GJ 2050       & BAG 22        & 06448+7153 & .8271 & 69.1$\ast$ & 0.7   & 545    & 7   &      &      & 800/110 \\
33142   & GJ 3412       & HEI 334       & 06541+6052 & .8271 & 186.3      & 0.8   & 187    & 3   &      &      & 800/110 \\
35457   & HD 56099      & HDS 1018      & 07192+5908 & .8272 & 16.8$\ast$ & 0.4   & 130    & 2   & 0.00 & 0.23 & 545/30  \\
38619   &               & HDS 1123      & 07545+6008 & .8272 & 178.5$\ast$& 0.5   & 688    & 6   &      &      & 800/110 \\
39261   & 53 Cam        & MCA 33        & 08017+6019 & .8243 & 305.3      & 0.7   & 90     & 2   & 1.41 & 0.02 & 545/30  \\
39402   &               & HDS 1149      & 08033+5251 & .8243 & 207.3      & 0.3   & 265    & 2   & 0.00 & 0.16 & 800/110 \\
    &               &               &            & .8272 & 208.1      & 0.6   & 266    & 3   &      &      & 800/110 \\
46199   & HD 81105      & HDS 1353      & 09252+4606 & .8244 & 153.7      & 0.3   & 361    & 2   & 2.48 & 0.03 & 600/30  \\
94679   & ADS 12239 AB  & STT 371       & 19159+2727 & .8231 & 160.4      & 0.3   & 881    & 2   & 0.27 & 0.20 & 545/30  \\
    &               &               &            & .8231 & 160.3      & 0.3   & 881    & 2   & 0.36 & 0.10 & 600/30  \\
    &               &               &            & .8231 & 160.1      & 0.3   & 881    & 2   & 0.27 & 0.07 & 800/110 \\
    &               &               &            & .8231 & 160.1      & 0.3   & 882    &     & 0.32 & 0.06 & 850/75  \\
    &               &               &            & .8258 & 160.5      & 0.3   & 882    & 2   &      &      & 545/30  \\
    &               &               &            & .8258 & 160.3      & 0.3   & 881    & 2   &      &      & 600/30  \\
    &               &               &            & .8259 & 160.1      & 0.3   & 883    & 2   &      &      & 800/110 \\
95178   & HD 183678     & HDS 2740      & 19218+7708 & .8261 & 2.0$\ast$  & 0.3   & 333    & 5   &      &      & 800/110 \\
95413   & CH Cyg        &               & 19246+5014 & .8151 & 24.1       & 2.1   & 43     & 2   & 2.03 & 0.04 & 545/30  \\
    &               &               &            & .8152 & 24.6       & 3.5   & 41     & 3   & 2.20 & 0.11 & 600/30  \\
95995   & GJ 762.1      & MCA 56        & 19311+5835 & .8150 & 75.0       & 0.3   & 110    & 2   & 0.29 & 0.03 & 600/30  \\
96339   & GJ 4114 A     & BAG 27        & 19351+0828 & .8232 & 3.8        & 0.3   & 284    & 2   & 0.17 & 0.06 & 800/110 \\
96656   & GJ 765.2      & MLR 224       & 19391+7625 & .8150 & 126.7      & 0.6   & 82     & 2   & 0.59 & 0.03 & 600/30  \\
97496   & ADS 12973 AB  & AGC  11       & 19490+1909 & .8149 & 28.4       & 0.4   & 74     & 2   & 0.39 & 0.05 & 545/30  \\
    &               &               &            & .8149 & 28.1       & 0.4   & 74     & 2   & 0.43 & 0.02 & 600/30  \\
    &               &               &            & .8149 & 28.3       & 0.6   & 74     & 2   & 0.38 & 0.06 & 800/110 \\
99874   & HR 7744       & MCA 60        & 20158+2749 & .8259 & 327.8      & 1.3   & 91     & 2   & 2.92 & 0.11 & 850/75  \\
101181  & HD 195397     & HDS 2932      & 20306+1349 & .8260 & 356.5$\ast$& 0.8   & 70     & 2   & 0.61 & 0.06 & 545/30  \\
102357  & GJ 804        & CAR 2         & 20444+1945 & .8260 & 139.5$\ast$& 1.9   & 86     & 3   &      &      & 800/110 \\
103502  &               & HDS 2989      & 20582+4011 & .8260 & 148.5$\ast$& 0.5   & 241    & 2   & 1.04 & 0.11 & 800/110 \\
103810  & ADS 14575 & STF 2751          & 21022+5640 & .8152 & 354.3      & 1.3   & 1617   & 6   &      &       & 545/30  \\
    &           &                   &            & .8152 & 353.6      & 1.3   & 1619   & 5   &      &       & 600/30  \\
    &           &                   &            & .8152 & 354.8      & 1.3   & 1622   & 5   &      &       & 800/110 \\
    &           &                   &            & .8234 & 354.4      & 1.3   & 1617   & 6   &      &       & 545/30  \\
    &           &                   &            & .8234 & 354.9      & 1.3   & 1619   & 6   &      &       & 600/30  \\
    &           &                   &            & .8234 & 354.8      & 1.3   & 1621   & 5   &      &       & 800/110 \\
    &           &                   &            & .8316 & 354.5      & 1.3   & 1615   & 5   &      &       & 545/30  \\
    &           &                   &            & .8317 & 354.7      & 1.3   & 1620   & 5   &      &       & 700/30  \\
    &           &                   &            & .8317 & 354.6      & 1.3   & 1622   & 5   &      &       & 800/110 \\
    &           &                   &            & .8342 & 355.2      & 1.3   & 1621   & 8   &      &       & 545/30  \\
    &           &                   &            & .8369 & 355.2      & 1.3   & 1618   & 6   &      &       & 700/30  \\
    &           &                   &            & .8369 & 356.0      & 1.3   & 1616   & 6   &      &       & 800/110 \\
104075  &  TV Equ   & HDS 3004          & 21051+0757 & .8233 & 3.6        & 0.6   & 252    & 3   & 3.87 & 0.10  & 545/30  \\
104565  & GJ 4182   & BAG 29            & 21109+2925 & .8233 & 210.3      & 0.4   & 126    & 2   & 0.43 & 0.05  & 800/110 \\
105187  & BD+65 1572 & HDS 3032         & 21185+6613 & .8317 & 143.9      & 0.3   & 737    & 3   &      &       & 800/110 \\
105259  & ADS 14864 Aa& BAG 9           & 21193+5837 & .8151 & 122.8      & 0.5   & 117    & 2   & 1.64 & 0.02  & 545/30  \\
    &           &                   &            & .8151 & 121.9      & 1.2   & 115    & 3   & 3.28 & 0.13  & 800/110 \\
    &           &                   &            & .8234 & 122.5      & 0.3   & 117    & 2   & 1.64 & 0.02  & 545/30  \\
    &           &                   &            & .8235 & 122.0      & 1.7   & 114    & 4   & 3.45 & 0.15  & 850/75 \\
105438  & ADS 14894 & STT 435           & 21214+0253 & .8318 & 236.5      & 0.3   & 679    & 2   & 0.55 & 0.07  & 800/110 \\
105947  & HD 204236 & HDS 3053          & 21274-0701 & .8152 & 127.5      & 0.3   & 195    & 2   & 1.47 & 0.03  & 545/30  \\
106059  & HD 204827 & HDS 3058          & 21290+5844 & .8261 & 181.3$\ast$& 0.9   & 93     & 2   & 0.88 & 0.04  & 545/30  \\
    &           &                   &            & .8261 & 181.8$\ast$& 0.7   & 93     & 2   & 0.92 & 0.03  & 700/30  \\
106886  & ADS 15184 Aa& MIU 2           & 21390+5729 & .8262 & 234.2$\ast$& 0.5   & 99     & 2   & 1.38 & 0.02  & 545/30  \\
    &           &                   &            & .8262 & 233.9$\ast$& 0.6   & 100    & 2   & 1.44 & 0.03  & 700/30  \\
108842  & HD 209421 & HDS 3129          & 22029+1547 & .8370 & 226.0      & 1.0   & 38     & 2   & 0.51 & 0.04  & 545/30  \\
109281  & HD 210211 & HDS 3145          & 22083+2409 & .8153 & 273.1      & 0.3   & 126    & 2   & 0.50 & 0.02  & 545/30  \\
109951  & HD 211276 & HDS 3158          & 22161-0705 & .8152 & 92.5       & 0.3   & 357    & 2   & 1.88 & 0.03  & 545/30  \\
112695  & HD 216027 & HDS 3241          & 22493+1517 & .8207 & 302.1      & 2.0   & 64     & 3   & 2.24 & 0.08  & 545/30  \\
    &           &                   &            & .8342 & 303.6      & 1.7   & 63     & 2   & 2.20 & 0.06  & 545/30  \\
112970  & HD 216606 & HDS 3247          & 22527+6759 & .8371 & 324.0      & 0.6   & 169    & 2   & 3.52 & 0.06  & 545/30  \\
113852  & HR 8778   & HDS 3285          & 23034+5834 & .8317 & 127.8      & 0.3   & 400    & 2   & 3.29 & 0.04  & 545/30  \\
114444  & HD 218793 & HDS 3302          & 23107+0947 & .8318 & 328.9      & 0.3   & 352    & 2   & 1.80 & 0.02  & 545/30  \\
114922  & GJ 893.4  & HDS 3316          & 23167+1937 & .8153 & 275.3      & 0.5   & 126    & 2   & 0.15 & 0.17  & 800/110 \\
    &           &                   &            & .8208 & 275.2      & 0.3   & 126    & 2   & 0.00 & 0.16  & 800/110 \\
114927  & BD+33 4679 & HDS 3315         & 23167+3441 & .8208 & 211.7      & 0.4   & 195    & 2   & 0.34 & 0.06  & 800/110 \\
115666  & ADS 16748 AB& HO 489            & 23260+2742 & .8154 & 223.0      & 0.4   & 521    & 3   & 0.00 & 0.30  & 545/30  \\
    &           &                   &            & .8154 & 223.3      & 0.4   & 519    & 3   & 0.37 & 0.03  & 600/30  \\
    &           &                   &            & .8154 & 223.1      & 0.4   & 517    & 3   & 0.76 & 0.02  & 800/110 \\
    &           &                   &            & .8207 & 222.8      & 0.4   & 517    & 3   & 0.00 & 0.30  & 545/30  \\
    &           &                   &            & .8207 & 223.3      & 0.4   & 517    & 3   & 0.34 & 0.05  & 600/30  \\
    &           &                   &            & .8207 & 223.5      & 0.4   & 518    & 3   & 0.79 & 0.02  & 800/110 \\
    &           &                   &            & .8207 & 223.3      & 0.4   & 517    & 3   & 0.91 & 0.04  & 850/75  \\
    &           &                   &            & .8372 & 222.7      & 0.4   & 520    & 3   & 0.00 & 0.30  & 545/30  \\
    &           &                   &            & .8372 & 223.3      & 0.4   & 519    & 3   & 0.64 & 0.03  & 700/30  \\
    &           &                   &            & .8372 & 223.3      & 0.4   & 517    & 3   & 0.76 & 0.02  & 800/110 \\
116259  & GJ 9830   & HDS 3356          & 23334+4251 & .8236 & 152.1      & 0.6   & 99     & 2   & 2.45 & 0.03  & 545/30  \\
116294  & HD 221630 & HDS 3357          & 23338-0508 & .8371 & 77.7       & 0.3   & 690    & 3   & 2.02 & 0.08  & 545/30  \\
116310  & ADS 16836 & BU 720            & 23340+3120 & .8236 & 97.3       & 0.3   & 550    & 2   & 0.38 & 0.04  & 600/30  \\
116810  &           & HDS 3363          & 23405+2959 & .8209 & 240.3      & 0.3   & 869    & 3   & 1.75 & 0.04  & 800/110 \\
118212  & GJ 913    &                   & 23587+4644 & .8210 & 74.2       & 0.9   & 62     & 2   & 1.36 & 0.03  & 850/75  \\
118287  & ADS 17151 & A 1498            & 23595+5441 & .8209 & 87.7       & 0.3   & 375    & 2   & 0.12 & 0.06  & 545/30  \\
    &           &                   &            & .8209 & 87.5       & 0.3   & 374    & 2   & 0.54 & 0.02  & 600/30  \\
    &           &                   &            & .8209 & 87.2       & 0.3   & 374    & 2   & 1.11 & 0.07  & 800/110 \\
    &           &                   &            & .8373 & 87.6       & 0.3   & 376    & 2   & 0.00 & 0.20  & 545/30  \\
    &           &                   &            & .8373 & 87.2       & 0.4   & 376    & 2   & 0.76 & 0.05  & 700/30  \\
\hline
\end{longtable}

\begin{table*}
\caption{Triple star measurements}
\label{tab2:Balega_n}
\bigskip
\begin{tabular}{l|l|l|c|c|c|c|c|c|c|c|c|l}
\hline
HIP      & Name/   & Discoverer & Coord. & Epoch   & Comp.&     $\theta$,\,deg       & $\sigma_{\theta}$, & $\rho$,   & $\sigma_{\rho}$ & $\Delta$m & $\sigma_{\Delta m}$ & $\lambda/\Delta\lambda$, \\
No.      & Catalog No. & designation & 2000.0 & 2004.0+ &          &                         & mas               & mas      &                 &           &                     & nm                    \\
\hline
5245    & HD 6639   & HDS 144  & 01071-0036  & .8155 & AB & 224.7 & 0.4 & 233 & 2 & 1.75 & 0.04 & 800/110 \\
        &           & BAG 12   &             &       & AC & 167.5 & 1.5 & 1223& 6 &      &      &         \\
101955  & GJ 795    & KUI 99   & 20396+0458  & .8232 & AB & 307.1 & 0.3 & 322 & 2 & 1.26 & 0.05 & 600/30  \\
    &           & BAG 14   &             &       & AC & 105.2 & 0.4 & 164 & 2 & 1.44 & 0.04 &         \\
    &           &          &             &       & BC & 119.8 & 0.3 & 479 & 2 & 0.18 & 0.07 &         \\
111805  & ADS 16138 & HO 295   & 22388+4419  & .8235 & AB & 153.8 & 0.3 & 314 & 2 &      &      & 545/30  \\
    &           & BAG 15   &             &       & AC & 162.2 &10.5 &  25 & 5 &      &      &         \\
    &           &          &             &       & BC & 153.1 & 1.0 & 290 & 5 &      &      &         \\
112170  & ADS 16214 & STT 476  & 22431+4710  & .8235 & AB & 119.6 & 0.3 & 492 & 2 & 0.35 & 0.05 & 545/30  \\
    &           &          &             &       & AC & 130.7 & 0.4 & 502 & 3 & 1.20 & 0.05 &         \\
    &           & HU 91    &             &       & BC & 209.1 & 1.5 & 97  & 3 & 0.86 & 0.05 &         \\
116384  & GJ 900    & MEL 9    & 23350+0136  & .8208 & AB & 335.7 & 0.3 & 610 & 2 & 2.56 & 0.06 & 800/110 \\
    &           &          &             &       & AC & 345.7 & 0.4 & 722 & 4 & 3.18 & 0.22 &         \\
116726  & ADS 16904 & A 643    & 23393+4543  & .8154 & AB & 138.9 & 0.3 & 250 & 2 & 0.11 & 0.02 & 545/30  \\
    &           &          &             &       & AC & 143.8 & 0.5 & 216 & 2 & 0.98 & 0.03 &         \\
    &           & CHR 149  &             &       & BC & 291.6 & 3.1 & 39  & 3 & 0.86 & 0.02 &         \\
\hline
\end{tabular}
\end{table*}

The differential measurements of magnitude differences $\Delta m$ between the
components were performed concurrently with the position measurements of the major
part of the studied stars. In Tables~\ref{tab1:Balega_n} and \ref{tab2:Balega_n} we give 142
$\Delta m$ values for binaries and 12 measurements for triples in different
bands. The uncertainty of the $\Delta m$ estimates varies from 0.02 to 0.37
magnitudes. Photometric measurements are more sensitive to seeing conditions
than astrometric; therefore, $\Delta m$ could not be derived with a
seeing worth than 2$''$. The main problem of differential speckle photometry
with the bad seeing is that the star partially falls outside the detector's
window (3$''$). The same difficulties arise during observations of wide pairs
with $\rho > 1''$. An example is GJ\,234, whose measurements of $\Delta m$
are marked with asterisks and probably overestimated because the frame window
cuts the speckle images.
\begin{figure}
\begin{center}
\includegraphics[width=7cm]{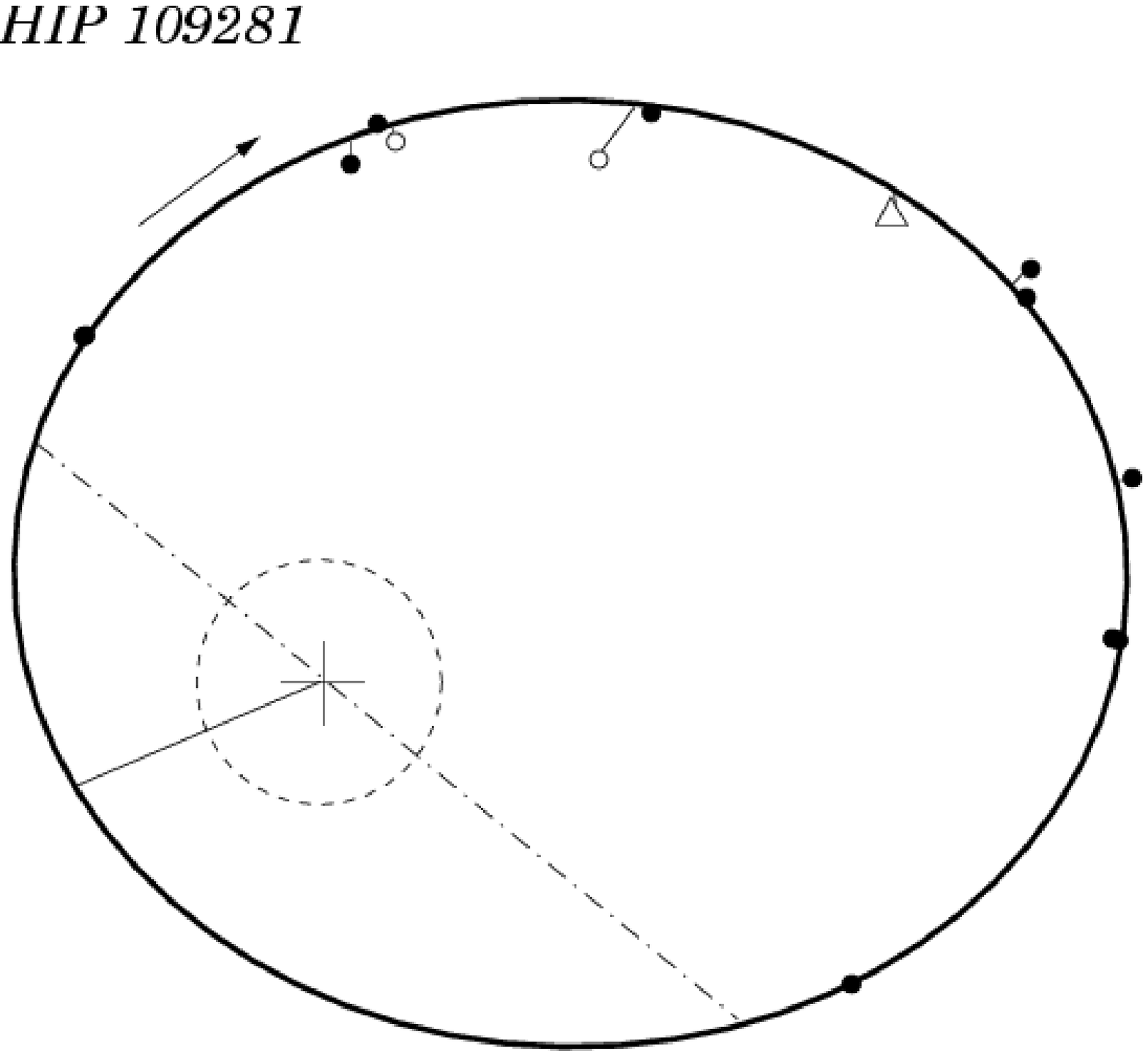}
\caption{Apparent ellipse representing the orbital
elements for HIP\,109281. The BTA speckle interferometric data are indicated
by filled circles, the speckle interferometric measurements performed by
Horch et al. are shown by open circles, and the Hipparcos measurement
is shown as an open triangle. Residual vectors for all measurements are
plotted, but in some cases they are smaller than the points themselves.
The orbital motion direction is indicated by an arrow. The solid line shows
the periastron position, while the dash-and-dot line represents the line of
nodes. The dashed circle around the position of the primary has an angular
radius of 0.02$''$ corresponding to the diffraction limit of the 6~m
telescope in the $V$ band. North is up and east is to the left.}
\label{fig1:Balega_n}
\end{center}
\end{figure}

\begin{figure}
\begin{center}
\includegraphics[width=6cm]{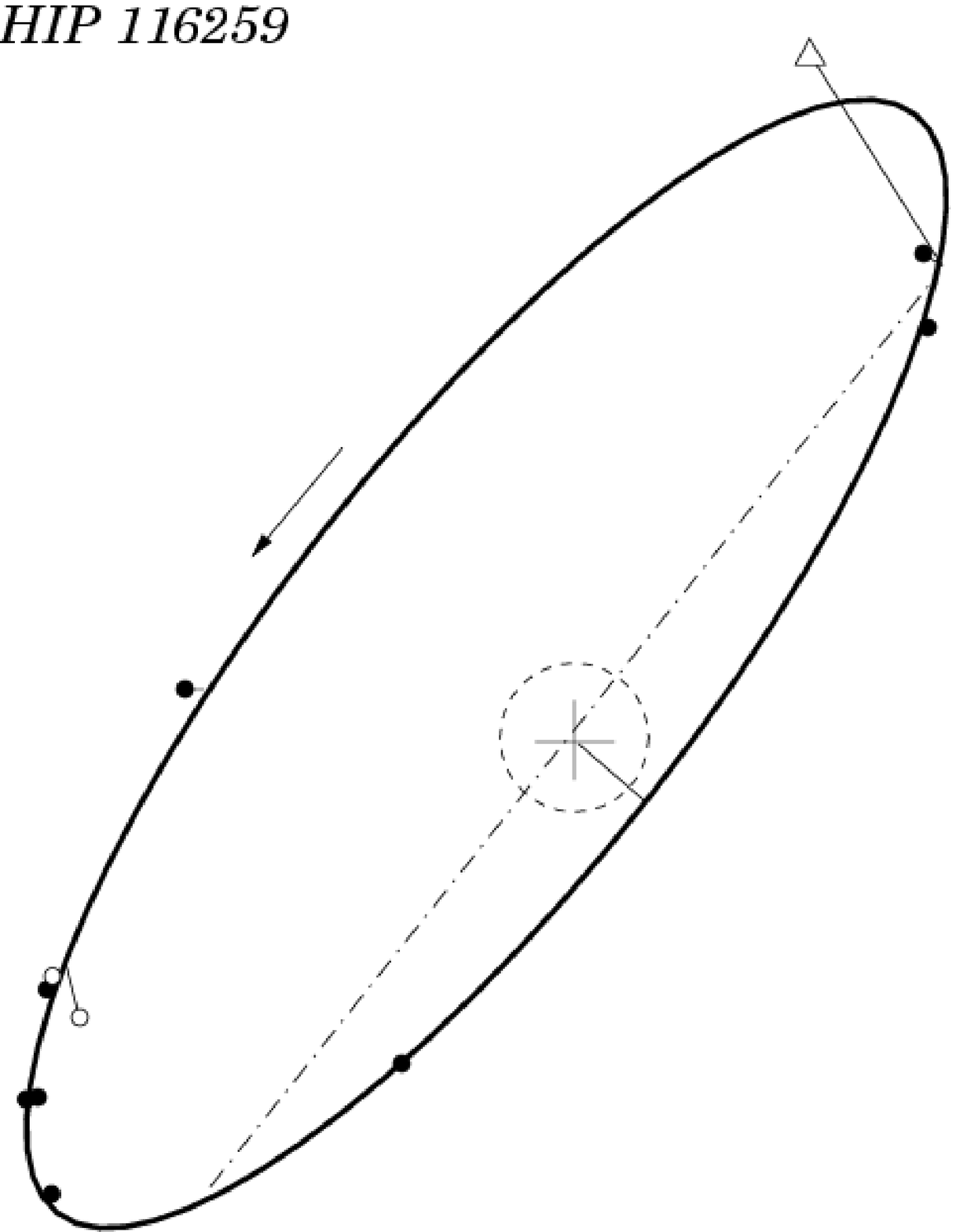}
 \caption{Apparent ellipse representing the orbital elements for HIP\,116259.}
\label{fig2:Balega_n}
\end{center}
\end{figure}

So far, 12 orbits for new Hipparcos binaries have been published based on
speckle interferometry with the BTA telescope (\cite{bag2005:Balega_n}; \cite{bag2006b:Balega_n}). Using
the 2004 observations and the newest 2006 measurements, we can derive orbital
parameters for two more Hipparcos binaries: HIP\,109281 and HIP\,116259. The
method of orbit computation is described in our previous paper
(\cite{bag2005:Balega_n}).  New relative orbits of the systems are plotted in
fig.~\ref{fig1:Balega_n} and \ref{fig2:Balega_n}, and short comments on the orbits are given in
the next section.

\section{COMMENTS ON INDIVIDUAL STARS}

{\bf HIP\,5245} (see Table~\ref{tab1:Balega_n}). The faint tertiary component (Bag\,12)
in this K0 system was first found in the K band with the BTA 6~m telescope
in 1999 (\cite{bag2002:Balega_n}). It turned out to be 3 magnitudes fainter than the
main component. A very weak sign of the component at a distance of 1.2$''$
from the A star was seen through the 800/110 nm filter in the 2004
observations. Its magnitude could not be estimated because of the noise in
the power spectrum.

{\bf HIP\,7338}. Nine interferometric observations and one Hipparcos
measurement of this pair of red dwarfs allowed us to confidently define half
of its relative motion ellipse. The resulting orbit, with a period of 23 yrs,
and the semimajor axis of a=0.20$''$ under $\pi_{Hp}$=28.7~mas gives a
mass-sum of the system of 0.7 $\mathcal{M}_{\odot}$. However, the
discrepancies between the measurements and the calculated positions are still
very high. It is possible that the first Hipparcos observation had a large
error, which may be explained by the faintness of the companion
(magnitude fainter than 13.5). It will probably take a few more years to define the
reliable orbit for the pair.

Our differential photometry carried out after 1999 shows that at 800 nm the
magnitude difference between the components is 2.2$\pm0.1^m$. Based on the
parallax value $\pi_{Hp}$=28.7~mas, the integral visible magnitude
$m_{V}$=10.64, and the color index $V$--$I$=1.35 in the Cousins system (\cite{eca:Balega_n}),
we derived the absolute $I$ magnitude of the secondary as $I_{B}$=9.0.
This corresponds to the M5 spectral type. The specified differential photometry
of the pair suggests a lower temperature of the secondary compared to
that proposed in the first paper of the series (\cite{bag2002:Balega_n}).

{\bf HIP\,39402}. This is another system of M dwarfs at a distance of 31~pc
from the Sun. Possibly, its orbital period is 26~yrs and the semimajor
axis is 0.28$''$. However, the scattering of speckle data is abnormally large
under this solution. We do not exclude that the significant deviations of the
measurements are caused by the presence of a third star in the system, as
mentioned in our 1999 observations (\cite{bag2004:Balega_n}).

{\bf HIP\,95413\,=\,CH Cyg}. A symbiotic system with the M6 giant main
component resolved for the first time. Presently, a generally accepted
model of the system does not exist; and there is no explanation for the
nature of the star's activity. Most researchers suggest a triple model
for CH\,Cyg: the M6III giant and a white dwarf form the inner orbit with a
period of 756 days, while the third component moves in the outer orbit with a
period of 14.5 yrs. The tertiary star could be a G-K dwarf or a giant
(\cite{hinkle:Balega_n}; \cite{skopal1995:Balega_n}; \cite{skopal2002:Balega_n}). However, the photometric and
spectroscopic variability of the star can also be satisfactorily interpreted
by a binary model (\cite{yamashita:Balega_n}; \cite{mikol:Balega_n}). Detailed analysis of the model
limitations from the speckle observations will be made in a separate paper.
Here we draw attention to only two important details. First, at a distance of
270~pc (\cite{viotti:Balega_n}), the discovered companion can only be connected with the
long-period orbit in the system. Second, the position angle of the pair
($\approx$25$\degr$) is almost perpendicular to the extended nebulosity
(position angle $\approx$165$\degr$) discovered in $UV$ continuum, [OIII] and
Balmer lines with the HST WFPC2 (\cite{eyers:Balega_n}).

{\bf HIP\,99874\,=\,23\,Vul\,=\,MCA\,60}. The main component of this pair is
a K3-type giant. A faint companion ($\Delta m$=2.92 in the $I$ band) is
moving with acceleration in the direction of the main star in a highly
inclined orbit. The pair was also observed with the BTA 6~m telescope in
2002. However, the results were never published because of bad weather
conditions during the observations: 2002.7980, $\theta$=326.8$\degr$,
$\rho$=156~mas. The quadrants of all speckle observations collected in the
4th Catalog of Interferometric Measurements of Binary Stars (\cite{wds:Balega_n})
have to be changed by 180$\degr$. The third star in the system, CHR\,94, has
never been detected in our observations.

{\bf HIP\,109281}. The elements of the interferometric orbit can now be
derived for this pair of evolved stars using 13 speckle measurements with the
BTA 6~m telescope and the WIYN 3.5 m telescope (\cite{horch:Balega_n}), and using the
first measurements by Hipparcos:

P\,=\,10.736$\pm$0.078 yrs,

T\,=\,1997.812$\pm$0.025,

e\,=\,0.518$\pm$0.010,

a\,=\,0.095$\pm$0.001$''$,

i\,=\,151.2$\pm$2.2$\degr$,

$\Omega$\,=\,50.4$\pm$3.2$\degr$,

$\omega$\,=\,292.5$\pm$3.2$\degr$.

Note that one new measurement with the 6~m telescope made in 2006 was used to
calculate the parameters (not included in Table~\ref{tab1:Balega_n}). The list of
position measurements, together with the deviations from calculated values,
is presented in Table~\ref{tab4:Balega_n}. The graphical presentation of the orbit
is shown in fig.~\ref{fig1:Balega_n}. With the Hipparcos parallax value
$\pi_{Hp}$=12.65$\pm$0.79~mas (\cite{eca:Balega_n}), the mass-sum of the binary is
\mbox{$\Sigma\mathcal{M}$=3.2$\pm$0.6 $\mathcal{M}_{\odot}$} (accuracy 19\%).

{\bf HIP\,111805\,=\,Bag\,15} (see Table~\ref{tab2:Balega_n}). A close companion of
the main star in the triple system is resolved marginally. Nevertheless,
despite the significant error of the measurement, we decided to include
its $\theta$ and $\rho$ values in Table~\ref{tab2:Balega_n}. The pair
(G2V+K1V, P=551 d) was discovered spectroscopically by Duquennoy (\cite{duq:Balega_n}) and later
resolved by speckle interferometry at the BTA telescope (\cite{bag2002:Balega_n}).

{\bf HIP\,116259}. The system with the G2 and K4 components (\cite{bag2002:Balega_n})
belongs to the old population of the Galaxy (\cite{nord:Balega_n}). 11 interferometric
measurements in the period from 1998 through 2006 allowed the determination
of the visible ellipse of the system's motion and the orbital parameters:

P\,=\,15.70$\pm$0.23 yrs,

T\,=\,2005.49$\pm$0.01,

e\,=\,0.536$\pm$0.007,

a\,=\,0.220$\pm$0.002$''$,

i\,=\,75.1$\pm$0.4$\degr$,

$\Omega$\,=\,141.5$\pm$0.3$\degr$,

$\omega$\,=\,89.5$\pm$0.8$\degr$.

The deviations of the measurements from the computed values are given in
Table~\ref{tab5:Balega_n}. The first measurement by Hipparcos was not taken into
account in the calculations. In addition to the data from Table~\ref{tab1:Balega_n},
we have made use of the newest measurement made at the BTA 6~m telescope in
2006. The ellipse of the interferometric orbit is shown in fig.~\ref{fig2:Balega_n}.
The total mass of HIP\,116259  is
$\Sigma\mathcal{M}$=1.56$\pm$0.18 $\mathcal{M}_{\odot}$ (accuracy 12\%)
under $\pi_{Hp}$=30.24~mas (\cite{eca:Balega_n}). As for all Hipparcos new binaries, the
parallax error plays a definitive role in the total error of the mass
estimate, but not the orbital elements. The spectroscopic orbit of the pair
(\cite{latham:Balega_n}) has similar characteristics.

{\bf HIP\,118212\,=\,GJ 913}. This nearby, $\pi_{Hp}$=58~mas (\cite{eca:Balega_n}),
M-type star was included in the program as a possible binary. It is one of
1561 stars in the Hipparcos Catalog marked with an X flag, which means that
only a stochastic solution for their astrometry was found. A part of these
stars can be non-single objects, while the other part can be explained by
the failure in data reduction. An attempt to improve the parallax of HIP\,118212 and
to define the character of its motion from the Hipparcos astrometry has
recently been made by Goldin and Makarov (\cite{goldin:Balega_n}). Using the results of
independent observation reductions by two consortia, FAST and NDAC, they
calculated a new parallax value for the star, $\pi_{Hp}$=67~mas, and defined
the orbit with a period of 885 days.

We first resolved a faint ($\Delta m$=1.4 in the $I$ band) close
($\rho$=62~mas) companion of HIP\,118212 with the BTA 6~m telescope in the
850/75~nm filter. Our measurement does not fit the calculated position on the
orbit of Goldin and Makarov (\cite{goldin:Balega_n}). The reason for this
discrepancy---a new companion or the wrong orbit---can be established in the immediate
future using new speckle observations of this presumably fast-moving pair.

\section{UNRESOLVED OBJECTS}
\begingroup
\begin{table*}
\begin{center}
\caption{Unresolved stars}
\label{tab3:Balega_n}
\bigskip
\begin{tabular}{l|l|c|c|l|c}
\hline
HIP     & Other       & Coord.     & Epoch       &  $\lambda/\Delta\lambda$,  & Note\\
No.     & catalog No.    & 2000.0  & 2004.0+  &  nm                          &      \\
\hline

916     & GJ 3012    & 00114+5821 & .8210 & 850/75   & X \\
1092    & GJ 3015 B  & 00136+8040 & .8210 & 800/110  & X \\
1295    &            & 00162+1952 & .8211 & 800/110  & X, S \\
1475    & GJ 15 A    & 00184+4401 & .8236 & 850/75   & S \\
1860    & GJ 1010 A  & 00235+7711 & .8211 & 800/110  & X, S \\
3362    & GJ 29.1    & 00428+3533 & .8320 & 800/110  & G \\
7765    & ADS 1307 B & 01399+1516 & .8238 & 800/110  & X, S \\
7981    & HR 493     & 01425+2016 & .8238 & 545/30   & O \\
16445   & GJ 143.3   & 03318+1419 & .8214 & 800/110  & X, S \\
19270   & ADS 2984 A & 04078+6220 & .8266 & 545/30   & \\
20222   &            & 04200+3629 & .8243 & 800/110  & X\\
31635   & GJ 239     & 06372+1734 & .8217 & 600/20   & S \\
36834   & GJ 277.1   & 07345+6256 & .8271 & 800/110  & S \\
97579   & HDS 2823   & 19500+3158 & .8232 & 545/30   & \\
97607   & HR 7554    & 19503+0754 & .8259 & 545/30   & \\
98538   & HD 189711  & 20011+0931 & .8259 & 800/110  & \\
101382  & GJ 793.1   & 20329+4154 & .8260 & 545/30   & \\
102851  & GJ 808.2   & 20502+2923 & .8233 & 600/20   & X, S \\
103256  & GJ 1259    & 20551+1311 & .8233 & 600/20   & \\
106886  & ADS 15184 C& 21390+5729 & .8262 & 545/30   & \\
        & ADS 15184 D&            & .8262 & 545/30   & \\
108467  & GJ 842.2   & 21584+7535 & .8235 & 800/110  & S \\
112460  & GJ 873 A   & 22468+4420 & .8319 & 800/110  & S \\
114941  & GJ 4323    & 23169+0542 & .8208 & 800/110  & S \\
117779  & GJ 910     & 23531+2901 & .8236 & 800/110  & X, S \\
117795  &            & 23533+5957 & .8210 & 800/110  & G \\
118310  & ADS 17154 A& 23598+0640 & .8318 & 700/30   & S \\
\hline
\end{tabular}
\end{center}
\end{table*}
\endgroup

\begingroup
\begin{table*}
\begin{center}
\caption[]{Position parameters and residuals of the measurements of the HIP\,109281}
\label{tab4:Balega_n}
\bigskip
\begin{tabular}{c|c|c|c|c|c}
\hline
Epoch     & $\theta$, & $\rho$, & $(O-C)_{\theta}$, & $(O-C)_{\rho}$, & Reference \\
      & deg     & mas    & deg             & mas            &        \\
\hline
1991.250  & 311.0    & 119  & -1.2             & -3               & \cite{eca:Balega_n}    \\
1998.774  & 32.5     & 68   &  0.3             &  1               & \cite{bag2002:Balega_n} \\
1999.741  & 356.7    & 86   &  0.1             & -4               & \cite{bag2002:Balega_n} \\
1999.821  & 354.4    & 93   &  -0.1            &  1               & \cite{bag2004:Balega_n} \\
1999.885  & 352.1    & 90   & -0.6             & -3               & \cite{horch:Balega_n}   \\
2000.764  & 333.1    & 97   &  0.1             & -10              & \cite{horch:Balega_n}  \\
2000.872  & 331.1    & 108  &  0.2             & -1               & \cite{bag2006a:Balega_n} \\
2002.736  & 301.4    & 131  &  0.2             &  4               & This paper   \\
2002.799  & 299.8    & 128  &  -0.5            &  1               & This paper   \\
2003.927  & 284.8    & 132  &  -0.2            &  3               & This paper   \\
2003.927  & 284.9    & 132  &  -0.1            &  3               & This paper   \\
2004.815  & 273.2    & 125  &  0.5             & -1               & This paper   \\
2004.815  & 273.1    & 126  &  0.4             &  0               & This paper   \\
2006.690  & 239.4    & 96   &  0.0             &  0               & This paper   \\
\hline
\end{tabular}
\end{center}
\end{table*}
\endgroup

A total of 26 objects were not resolved in the course of the observations.
They are listed in Table~\ref{tab3:Balega_n}. Due to the marginal weather conditions,
some of the binaries with a limiting magnitude difference for speckle
interferometry (around 3.5 magnitudes) were not resolved. One example of such
a system is HIP\,97579, with a remote companion ($\rho$=687~mas,
$\Delta m$=3.46) given in the Hipparcos Catalog. The power spectrum of this
pair can be traced up to the limiting frequency of the telescope, but its
noisiness could be the reason why the secondary was not detected.

Another unresolved star HIP19270=ADS2984A is a southern component
of the wide visual pair. It was observed under poor seeing conditions as a
reference source for its northern neighbor ADS\,2984\,B, which is known as
SZ\,Cam---a distant occultation binary with early-type massive components.

One more unresolved star, HIP\,97607\,=\,CHR\,89 of B2IVe spectral type,
appeared in the lists of new binaries after speckle observations with the
CFHT 3.6~m telescope in 1985 (\cite{mcalister1987a:Balega_n}). Another observation of
the star was obtained in the same year by the same authors using the MAYAL
3.8 m telescope at Kitt Peak (\cite{mcalister1987b:Balega_n}). However, neither the
Hipparcos observations nor the following BTA speckle interferometry confirmed
the duplicity. Despite the poor seeing conditions, our observations in 2004
allowed us to study the power spectrum up to the highest spatial frequencies.
We do not exclude the possibility that the pair will prove to be a
short-period system (P$\approx$50 yr), which is presently unresolved.

The history of speckle observations of HIP\,98538 = CHR\,118 is similar to
the history of the previous star CHR\,89. After the only measurement in
1985 (\cite{lu:Balega_n}), the companion has never been observed again. Note that the
confirmation of the binary nature of the CH star CHR\,118 is of great
importance for the explanation of the properties of this rare stellar type.

Following the accurate orbit of the spectroscopic and interferometric
binary HIP\,101382\,=\,HD\,195987 = GJ~793.1 with a period of 57.3 days
(\cite{torres:Balega_n}), the separation between the components in the period of the BTA
observations was only 9~mas. That explains our negative result because such
a separation is smaller than the diffraction limit of the 6~m aperture.
Earlier, Blazit et al. (\cite{blazit:Balega_n}) reported the speckle interferometric
resolution of the system with the CFHT 3.6~m telescope in 1985:
$\theta$=170$\degr$, $\rho$=30~mas. The ephemeris separation value in the
period of their observation was also equal to 9~mas; therefore, the binary
could not be resolved at CFHT, which has a diffraction limit of $\approx$40~mas.

In the spectroscopic binary system HIP\,103256\,= GJ\,1259, the companion's
mass is 7 times lower than the mass of the main K3V star (\cite{halbw:Balega_n}).
The luminosities of a K3V star and a late red dwarf differ in the visible by
7--8 magnitudes, ruling out the possibility for speckle resolution of the
components.

The system ADS\,15184\,C,D (HIP\,106886) is a member of the OB star complex.
The stars were included in the program as reference sources for the triple
star ADS\,15184\,A\,=\,MIU\,2, which includes both a spectroscopic pair and a
remote O companion (\cite{burkh:Balega_n}). Mason et al. (\cite{mason1998:Balega_n}) observed the
C and D components of ADS\,15184 earlier and could not detect a sign of their
multiplicity. It should also be taken into account that our observations were
carried out under poor seeing conditions.

The nearby ($\pi_{Hp}$=39~mas) K5V star HIP\,118310 = ADS\,17154\,Aa\,=\,Bag\,31
was first resolved with the BTA 6~m telescope in 2001 with $\rho\approx$0.2$''$.
Three years later, the secondary was not detected despite the fact that the power
spectrum was accumulated to the limiting frequencies. We conclude from this
that the pair can show a fast orbital motion with a period of $\approx$10
years.

In the last column of Table~\ref{tab3:Balega_n}, we present flags for the Hipparcos
Catalog ``problem'' stars with the following astrometric solutions: G,
motion with acceleration, X, stochastic solution for the photocenter motion,
S, possible non-single system. Flag O stands for HIP\,7981 with the computed
Hipparcos astrometric orbit. Following this orbit, the binary is too close to
be resolved with the 6~m telescope. As it follows from our earlier
observations, up to 30--35\% of the Hipparcos ``problem'' stars could be
resolved using the speckle interferometry at the BTA 6~m telescope
(\cite{bag2006a:Balega_n}). In the 2004 observations, only one out of six new stars in
this category was resolved for the first time (HIP\,118212). Other Hipparcos
``problem'' stars (HIP\,916, 1092, 1475, 1860, 3362, 7765, 7981, 16445,
31635, 36834, 108467, 112460, 117795) still remain unresolved.

\begin{table*}
\begin{center}
\caption[]{Position parameters and residuals of the measurements of the HIP\,116259}
\label{tab5:Balega_n}
\bigskip
\begin{tabular}{c|c|c|c|c|c}
\hline
Epoch     & $\theta$, & $\rho$, & $(O-C)_{\theta}$, & $(O-C)_{\rho}$, & Reference \\
      & deg     & mas    & deg             & mas            &        \\
\hline
1991.25   & 341.0    & 195  & 18.9             & 34     & \cite{eca:Balega_n}    \\
1998.775  & 83.0     & 105  & 0.4              & 6      & \cite{bag2002:Balega_n} \\
2000.617  & 119.6    & 153  & 6.2              & 4      & \cite{horch:Balega_n}   \\
2000.759  & 114.7    & 154  & -0.2             & 1      & \cite{horch:Balega_n}   \\
2000.865  & 115.6    & 157  & -0.4             & 1      & \cite{bag2006a:Balega_n} \\
2000.873  & 115.6    & 157  & -0.5             & 1      & \cite{bag2006a:Balega_n} \\
2001.761  & 123.5    & 177  & -0.3             & 0      & \cite{bag2006a:Balega_n} \\
2001.761  & 123.8    & 174  & 0.0              & -3     & \cite{bag2006a:Balega_n} \\
2002.796  & 130.9    & 185  & -0.3             & -3     & This paper   \\
2004.824  & 152.1    & 99   & 0.2              & 0      & This paper   \\
2006.690  & 319.3    & 146  & -0.1             & 1      & This paper   \\
2006.946  & 321.9    & 161  & -0.1             & 0      & This paper   \\
\hline
\end{tabular}
\end{center}
\end{table*}

\section{CONCLUSION}

Speckle interferometric observations of 110 binary and multiple stars were
taken in 2004 October at the BTA 6~m telescope with the diffraction
resolution of the aperture: 19~mas in the 545/30~nm filter, 21~mas in the
600/30~nm filter, 28~mas in the 800/110 nm filter, and 29~mas in the 850/75~nm
 filter. Most of the objects in the program are nearby late-type dwarfs.
About half of them are new pairs discovered by the Hipparcos astrometric
satellite.

197 measurements of position angles and distances between the components of
multiple systems have been collected in Tables~\ref{tab1:Balega_n} and \ref{tab2:Balega_n}. The
errors of the measurements range from 0.3$\degr$ to 3.1$\degr$ for the
position angle and from 2 to 8~mas for the angular separation. The closest
among the resolved pairs is $\theta^{1}$\,Ori\,C with a separation between the
components of 23~mas, corresponding to 83\% of the limiting resolution. The
widest observed pair is ADS\,14575 ($\rho$=1.6$''$), which is a standard star
for the binary star speckle interferometry. In a separate table, we give a
list of 26 stars that remained unresolved in 2004.

In this paper we presented 154 measurements of the brightness difference
between the components of binary and multiple stars. In the last few years,
this has become the standard procedure in stellar speckle interferometry and is
significant for deriving the physical properties of studied stars.

The symbiotic system CH\,Cyg and the pair of red dwarfs HIP\,118212 were resolved
for the first time. The observations of the binary system CH\,Cyg are of
particular importance because up to now, there has been no satisfactory model for this
nearby symbiotic star, and the nature of its activity remains unclear. Nor has
it been determined whether the carbon-oxygen white dwarf in the system
is able to accumulate mass due to accretion from its cool companion until it
approaches the Chandrasekhar limit and becomes a supernova SN Ia progenitor.

Using BTA speckle interferometry, we obtained first orbits for two binaries: the
CN giant system HD\,210211\,=\,HIP\,109281 with a period of 10.7~yrs, and the
G2V-K4V pair GJ\,9830\,=\,HIP\,116259 with a period of 15.7~yrs. Their
orbital elements and (O--C) deviations from the predicted positions are
presented in the paper.

\end{document}